\documentclass[aps,prb,twocolumn,groupedaddress]{revtex4}
\usepackage{graphicx}

\begin{document}
\newcommand{\BSCCO}{Bi$_2$Sr$_2$CaCu$_2$O$_{8+\delta}$}

\title{Homogenous nodal superconductivity coexisting with inhomogeneous charge order
 in strongly underdoped \BSCCO}


\author{K. McElroy}
\affiliation{Physics Department, University of California,
Berkeley, CA 94720 USA\\LASSP, Department of Physics, Cornell
University, Ithaca, NY 14850 USA}

\affiliation{}
\author{D.-H. Lee}
\affiliation{Physics Department, University of California,
Berkeley, CA 94720 USA\\Material Sciences Division, Lawrence
Berkeley National Lab., Berkeley, CA 94720  USA}
\author{J. E. Hoffman}
\affiliation{Department of Applied Physics, Stanford University,
Stanford, CA 94305, USA}
\author{K. M Lang}
\affiliation{Department of Physics, Colorado College, CO 80305,
USA}

\author{E. W. Hudson}
\affiliation{Department of Physics, MIT, Cambridge MA 02139, USA}
\author{H. Eisaki}
\affiliation{AIST, 1-1-1 Central 2, Umezono, Tsukuba, Ibaraki,
305-8568 Japan}
\author{S. Uchida}
\affiliation{Department of Physics, University of Tokyo, Tokyo,
113-8656 Japan}
\author{J. Lee}
\affiliation{LASSP, Department of Physics, Cornell University,
Ithaca, NY 14850 USA}
\author{J.C. Davis}
\affiliation{LASSP, Department of Physics, Cornell University,
Ithaca, NY 14850 USA}
 \email[]{jcdavis@ccmr.cornell.edu}
\homepage[]{http://people.ccmr.cornell.edu/~jcdavis}


\date{\today}

\begin{abstract}
We use novel STM techniques in concert to study the doping
dependence of electronic structure in \BSCCO. At all dopings, the
low energy states are relatively homogenous except for dispersive
density-of-states modulations whose properties are used to
elucidate the momentum-space characteristics of quasi-particles.
The superconductive coherence-peaks, ubiquitous in near-optimal
tunneling spectra, are destroyed with strong underdoping. A new
spectral type, likely characteristic of the zero temperature
pseudogap regime, appears in these samples. Exclusively in regions
exhibiting this new spectrum, we find quasi periodic modulations
in dI/dV as well as in topography, with $\vec{Q}=(\pm
2\pi/4.5a_0,0)$ and $(0,\pm 2\pi/4.5a_0)\pm 15\%$.  This is
consistent with the existence of a local charge density modulation
at these wave vectors. Surprisingly, this state coexists
harmoniously with the low energy nodal quasi-particles. We discuss
the relevance of these findings to the cuprate phase diagram and
to the relationship between the pseudogap and superconductivity.
\end{abstract}

\pacs{71.18.+y}

\maketitle High temperature superconductivity emerges in the
cuprates when localized electrons in the CuO$_2$ plane become
itinerant due to hole-doping.\cite{orenstein00} A schematic
diagram of the dependence of their electronic/magnetic
characteristics on temperature (T), and doped-hole density ($p$),
is shown in

Fig. 1 (adapted from Ref. 2). This is referred to colloquially as
the cuprate 'phase-diagram' although true electronic phases
(antiferromagnetic insulator (AFI), superconductor (SC) and metal
(M)) have been identified in only three of its regions. Among the
other poorly understood and sometimes overlapping regions are
those characterized by a 'pseudogap'
(PG),\cite{timusk99,norman03,orenstein00} as a disordered magnetic
'spin glass' (SG) with incommensurate spin fluctuations
,\cite{yamada98,niedermayer98} and as a non-Fermi liquid
(NFL).\cite{varma02,norman03} Although no widely accepted theory
exists for these phenomena, research into the underdoped cuprates
($p<0.16$) is primarily motivated by two (apparently) distinct
points of view. In the first, another electronic phase (ordered
state) competes with the superconductivity. This state is thought
to drive the superconducting critical temperature T$_c$ to zero as
it strengthens with reduced doping, and to disappear near maximum
T$_c$ at a hidden quantum critical point (X in Fig. 1). Within
this picture, the pseudogap is primarily a property of an unknown
electronic ordered state. The second picture is one in which,
because of low superfluid density at low doping in two dimensions,
superconductivity is destroyed by order-parameter phase and/or
amplitude fluctuations and, as superfluid density rises with
increased hole-doping, the T$_c$ rises. From this perspective, the
pseudogap characterizes a region above T$_c$ exhibiting gaps in
the spin and charge spectrum due to the existence of singlet
pairs, even though the material is not a superconductor. Here we
report on doping-dependent STM studies of \BSCCO (Bi-2212),
designed to explore these two models.

A serious complication for experimental exploration of these ideas
is the fact that real-space ($\vec{r}$-space) probes generally
detect nanoscale spatial heterogeneity in the electronic/magnetic
structure of underdoped cuprates. For example, when $0.03<p<0.14$,
muon spin rotation ($\mu$SR) studies indicate the existence of a
disordered magnetic 'spin glass' in La$_{2-x}$Sr$_x$CuO$_4$
(La-214) and oxygen doped LaCuO$_{4+\delta}$,\cite{uemura02} and
in \BSCCO (Bi-2212).\cite{panagopoulos02} Scanning tunneling
microscopy (STM) based local-density-of-states (LDOS) imaging
reveals nanoscale electronic structure variations in
(Bi-2212)\cite{cren01,pan01,howald01,lang02,matsuda03} and in
Na$_x$Ca$_{2-x}$CuO$_2$Cl$_2$.\cite{koshaka03} Nuclear magnetic
resonance (NMR) points to strong nanoscale carrier density
disorder with variations in local $p$ of at least $25\%$ of mean
carrier density in both underdoped La-214\cite{hasse02,singer02}
and underdoped Bi-2212.\cite{loram02} For a wide variety of
underdoped cuprates, scaling analyses of penetration depth
measurements reveal finite size effects consistent with nanoscale
heterogeneity in the superfluid density.\cite{schneider03} These
results provide abundant independent evidence that spin and charge
degrees of freedom are heterogeneous at the nanoscale in many
underdoped cuprates. It has not yet been possible to determine if
this heterogeneity is a sample-specific and extrinsic effect due
to crystal, dopant, or chemical disorder, or is an intrinsic
effect of the cuprate electronic structure. Nor have its
implications for the phase diagram been considered widely.

A complementary description of electronic structure to that in
$\vec{r}$-space is in momentum space ($\vec{k}$-space), accessible
for cuprates via angle resolved photoemission (ARPES), and optical
techniques. ARPES reveals that, at optimal $p$ in the
superconducting phase, the Fermi-surface (FS) of hole-doped
cuprates is gapped by an anisotropic energy gap $\Delta(\vec{k})$
with four nodes, and, below T$_c$, quasiparticles exist everywhere
along the normal-state Fermi
surface.\cite{damascelli03,campuzano02,johnson01} At a fixed low
temperature, those quasi-particle states with $\vec{k}=
(\pm\pi/a_0,0)$ and $(0,\pm\pi/a_0)$ near the 1st Brillouin
zone-face, degrade rapidly in coherence with reduced doping until
they have become incoherent at
$p<0.10$.\cite{damascelli03,campuzano02,fedorov99,loesser97} This
is a mysterious phenomenon, since it is closely correlated with
superfluid density\cite{feng00} but, within conventional
superconductivity theories, should not be related to it. By
contrast, states on the 'Fermi-arc' (FA)\cite{norman98} nearby the
nodes retain their coherence down to the lowest dopings
studied.\cite{damascelli03,yoshida03,ronning03} Transient grating
optical spectroscopy studies of non-equilibrium quasi-particles in
underdoped YBa$_2$Cu$_3$O$_{6.5}$\cite{gedik03} also find
lifetimes for antinodal excitations that are orders of magnitude
shorter than those of the nodal quasi-particles. Thus, the
electronic structure of underdoped cuprates also appears to be
heterogeneous in $\vec{k}$-space, in the sense that states
proximate to the gap-nodes (nodal) have quite different
characteristics and evolution with doping, than those near the
zone-face (antinodal). No generally accepted explanation exists
for either these $\vec{k}$-space phenomena or their relationship
to the phase diagram.

An electronic structure like this, which is heterogeneous in both
$\vec{r}$-space and  $\vec{k}$-space, presents special challenges
to conventional experimental techniques. Nevertheless for the
cuprates, it is critical to understand the relationship between
these two electronic structure representations and how it evolves
with doping. Here, by using several innovative STM techniques in
concert, we have carried out the first comprehensive study of the
doping dependence of Bi-2212 superconducting electronic structure
in $\vec{r}$-space and determined its relationship to the
electronic structure in  $\vec{k}$-space. Our first technique is
local density of states ($LDOS$) mapping which consists of
measurement of the tip-sample differential tunneling conductance
$g(\vec{r},V)\equiv\frac{dI}{dV}|_{\vec{r},V}$ at each spatial
location and at each sample bias voltage V. Since
$LDOS(\vec{r},E=eV)\propto g(\vec{r},E=eV)$, an energy-resolved
$\vec{r}$-space image of the electronic structure is attained.
Second, from $g(\vec{r},E)$ a dataset, the magnitude of the
energy-gap in the density of states $\Delta(\vec{r})$ can be
determined in a process called a gap-map. Here
$\Delta(\vec{r})\equiv
\frac{\Delta(\vec{r})_{+}-\Delta(\vec{r})_{+}}{2}$ and is the
energy of the first maximum in $LDOS$ above(below) the Fermi level
(neglecting impurity states).\cite{lang02} A final technique,
recently introduced to cuprate studies,\cite{hoffman02v} is
Fourier transform scanning tunneling spectroscopy (FT-STS). Here,
the $\vec{q}$-vectors of spatial modulations in $g(\vec{r},E)$ are
determined from the locations of peaks in $g(\vec{q},E)$, the
Fourier transform magnitude of $g(\vec{r},E)$. This technique has
proven valuable by virtue of its exceptional
capability\cite{hoffman02q,mcelroy03,vershinin04} to relate the
nanoscale $\vec{r}$-space electronic structure to that in
$\vec{k}$-space.

To clarify these techniques, we show examples of each measurement
type in Fig.2. All are derived from a single $g(\vec{r},E)$ data
set measured with $~1.8$ \AA resolution in $256^2$ pixels on a 50
nm -square field of view (FOV).  Figure 2{\bf A} is the gapmap
$\Delta(\vec{r})$ from this $g(\vec{r},E)$ using a color scale
that spans $20 {\mathrm meV}<\Delta(\vec{r})<70$ meV . The
superposed inset shows the resolution of a topographic image taken
precisely where this $g(\vec{r},E)$ was measured. An example of
unprocessed $g(\vec{r},E)$ typical of strongly underdoped samples,
is shown in Fig.2{\bf B}. These spectra were measured along the
red line in Fig 2{\bf A}. From these data, one can see dramatic
changes during which the coherence peaks in the spectrum disappear
and are replaced by to a distinctly different type of spectrum.
Such evolutions are ubiquitous in strongly underdoped Bi-2212
samples. Fig.2{\bf C} is the measured $g(\vec{r},E=-12$ meV) in
the FOV of 2{\bf A}. It exhibits a complex pattern of very weak
$LDOS$ modulations which are relatively homogenous in the sense
that RMS-deviations from spatially averaged value of
$g(\vec{r},E)$  are only ~10\%. Fig.2{\bf D} is the
$g(\vec{q},E=-12$ meV) calculated from Fig.2{\bf C} and it reveals
that the $LDOS$-modulations in Fig.2{\bf C} were made up of only a
small set of well-defined
$\vec{q}$-vectors.\cite{hoffman02q,mcelroy03}

We apply these techniques in concert to study the
doping-dependence of electronic structure in a series of Bi-2212
samples. They are all single crystals grown by the floating zone
method. Doping is controlled by oxygen depletion so that no other
elemental impurities are introduced.  Each is cleaved in cryogenic
ultra-high vacuum before immediate insertion in to the STM head.
If its BiO surface is flat and free of nanoscale debris, each
sample is usually studied for several months, typically in a ~50nm
square field of view (FOV). More than $10^6$  spectra were
acquired for the studies reported here.

In Fig.3 we show 50nm-square gapmaps measured on samples with five
different dopings. Identical color scales representing 20
meV$<\Delta(\vec{r})<70$ meV are used for all images. The local
hole concentration is impossible to determine directly, but we
estimate that the bulk dopings were approximately 3{\bf
A}($0.19\pm0.01$), 3{\bf B}($0.18\pm0.01$), 3{\bf
C}($0.15\pm0.01$), 3{\bf D}($0.13\pm0.01$), 3{\bf
E}($0.11\pm0.01$). Near optimal doping (Fig 3{\bf A},{\bf B}) the
gapmaps are heterogeneous but nonetheless the vast majority of
tunneling spectra are manifestly those of a superconductor (see
below). However, at the lowest dopings and for gap values
exceeding approximately 65 meV, there are very many spectra where
$\Delta$ actually becomes ill defined because coherence peaks do
not exist at the gap edge (see for example Fig.3{\bf F}, spectrum
6). We represent these spectra by black in the gapmap, since they
are almost identical to each other and appear to be the limiting
class of spectra at our lowest dopings.

The spatially averaged value of $\Delta(\vec{r})$ for each
crystal, $\bar{\Delta}$, and its full width at half maximum,
$\sigma$, are: 3{\bf A}($\bar{\Delta}=33\pm1$ meV, $\sigma$=7
meV), 3{\bf B}($\bar{\Delta}=36\pm1$ meV, $\sigma$=8 meV), 3{\bf
C}($\bar{\Delta}=43\pm1$ meV, $\sigma$=9 meV), 3{\bf D}( $48\pm1$
meV, $\sigma$=10 meV), and 3{\bf B}($\bar{\Delta}>62$ meV but with
$\sigma$ ill defined). As doping is reduced, $\bar{\Delta}$ grows
steadily consistent with other spectroscopic techniques, such as
ARPES and break-junction
tunneling,\cite{damascelli03,campuzano02,miyakawa98} which average
over the heterogeneous nanoscale phenomena. This observation is
very important because it demonstrates that our Bi-2212 surfaces
evolve with doping in an electronically equivalent fashion to
those studied by the other techniques, and that we are probing the
low temperature state of the underdoped pseudogap regime.

In Fig.3{\bf F} we show a series of the 'gap-averaged' spectra.
Each is the average spectrum of all regions exhibiting a given
local gap value (from the single ~50nm FOV of Fig.2{\bf C}). They
are color-coded so that each gap-averaged spectrum can be
associated with regions of the same colour in all gapmaps (Fig.'s
3{\bf A-E}). The spectra are labeled from 1-6, with numbers 1
through 4 providing clear examples of what we refer to as
coherence peaks at the gap edge (indicated by the arrows). These
gap-averaged spectra are consistent with data reported previously
by from gapmap studies by Matsuda {\it et al}.\cite{matsuda03}
Here, from our doping dependence study, we can report that this
set of gap-averaged spectra is almost identical for all dopings.
The dramatic changes with doping seen in $\Delta(\vec{r})$ (Fig.3)
occur because the probability of observing a given type of
spectrum in Fig.3{\bf F} evolves rapidly with doping. For example,
the gap-averaged spectrum labeled as 1 in Fig.3{\bf F} has a 30\%
probability of occurring in gapmap 3{\bf A}, 25\% in 3{\bf B}, 5\%
in 3{\bf C}, less than 1\% in 3{\bf D}, and 0\% in 3{\bf E}. The
spectrum labeled 6 has a 0\% probability of occurring in 3{\bf A},
0.1\% in 3{\bf B}, 1\% in 3{\bf C}, 8\% in 3{\bf D} and $>55$\%
probability in 3{\bf E}. The evolution of these gapmaps (Fig.3)
with falling doping, from domination by heterogeneous but
predominantly superconducting spectral characteristics (Fig.3{\bf
A},{\bf B}) to domination by spectra of a very different type
(Fig.3{\bf E}) is striking.

Despite the intense changes with doping in the gapmaps (whose
information content is, by definition, derived from the coherence
peaks at $E=\Delta(\vec{r})$), the $LDOS$ at energies below about
0.5$\bar{\Delta}$ remains relatively homogenous for all dopings
studied. Figure 3{\bf F} reveals this low-energy $LDOS$
homogeneity because, independent of gap value, the $g(E)$ below
$\approx25$ meV are almost the same everywhere and for all
spectra. These low energy $LDOS$ do, however, exhibit numerous
weak, incommensurate, energy-dispersive, spatial
$LDOS$-modulations with long correlation lengths (for example
Fig.2{\bf C}). We focus on the doping dependence of these low
energy $g(\vec{r},E)$ data by applying the FT-STS technique.
Figure

4{\bf A-C} shows measured $g(\vec{q},E)$ for the three
$g(\vec{r},E)$ datasets used to generate Fig.3{\bf A}, 3{\bf D},
and 3{\bf E}. Each sub-panel is the measured $g(\vec{q},E)$ at the
labeled energy, with the reciprocal space locations of the Bi (or
Cu) atoms $\vec{q} = (\pm2\pi/a_0,0)$ and $(0,\pm2\pi/a_0)$,
appearing as the four dark spots at the corners of a square.  It
is obvious that multiple sets of dispersive $LDOS$-modulations
exist at all three dopings, but each exhibits different
trajectories as a function of $E$ for different $p$.

Analysis of these low energy $LDOS$ modulations requires a model
for their relationship to states in  $\vec{k}$-space. We apply the
"octet model" of quasiparticle
interference\cite{hoffman02q,mcelroy03} which is predicated on a
Bi-2212 superconducting band-structure exhibiting four sets of
'banana'-shaped contours of constant quasiparticle-energy
surrounding the gap nodes.\cite{damascelli03} Because of the
quasiparticle density of states at $E$ is
\begin{equation}
n(E)\propto \oint_{E(k)=\omega} \frac{1}{\nabla_{k}E(\vec{k})}dk
\end{equation}
while each 'banana' exhibits its largest $|
\nabla_{k}E(\vec{k})|^{-1}$ near its two ends, the primary
contributions to come from the octet of momentum-space regions at
the ends of each 'banana' $\vec{k}_j$; j=1,2,..8 (Fig.5{\bf A}).
Mixing of quasiparticle states in the octet by disorder scattering
produces quasiparticle interference patterns which are manifest as
spatial $LDOS$-modulations. The intensity of such scattering
induced modulations is primarily governed by joint density of
states (among other factors). The wavevectors of the most intense
LDOS-modulations are then determined by all possible pairs of
points in the octet $\vec{k}_j$. Sixteen distinct $+\vec{q}$ and
$-\vec{q}$ pairs should be detectable at each non-zero energy by
FT-STS. From them, the energy dependence of the octet locations
$\vec{k}_j(E)$ can be determined and associated with a 'locus of
scattering' $\vec{k}_s(E)$. Comprehensive internally-consistent
agreement between Bi-2212 STM data and this model is achieved near
optimal doping.\cite{mcelroy03} Until this work, nothing was known
about its utility for strongly underdoped cuprates.

Theoretical analyses beyond the simple octet
model\cite{bishop02,byers93,zhang03,wang03,zhu01,capriotti03,zhu03,chen03yeh,peregbarnea03,bena03}
capture many elements of our previously reported $g(\vec{q},E)$
data, but no resolution of the exact source, strength, or type of
scattering has yet been achieved. Nevertheless, the existence of
numerous sets of long-correlation length, dispersive, $LDOS$
modulations, all of which are self-consistent with a single
$\Delta(\vec{k})$ for both filled and empty states, is indicative
of good Bogoliubov-like quasi-particles. Since the
$LDOS$-modulations can be associated consistently with a 'locus of
scattering' $\vec{k}_s(E)$ via the octet model, we analyze our
observations within this model using the  $\vec{q}$-vector
designations shown in Fig.5{\bf A}.

Figure 5{\bf B} shows the measured length of $\vec{q}_1$,
$\vec{q}_5$ and $\vec{q}_7$ as a function of energy for the three
datasets in Fig.4. Figure 5{\bf C} shows the locus of scattering
calculated for these three using:
\begin{eqnarray}
\vec{q}_1=(2k_x,0) ; \vec{q}_5=(0,2k_y) ;
\vec{q}_7=(k_x-k_y,k_y-k_x)\\
\vec{k}_s=(\pm k_x(E),\pm k_y(E));\vec{k}_s=(\pm k_y(E),\pm
k_x(E))
\end{eqnarray}
The $\vec{k}_s(E)$ determined by this technique differs only
slightly between dopings. even though the actual $g(\vec{r},E)$
for different dopings are quite different at any given energy.
These three $\vec{k}_s(E)$ are each the same for filled and empty
sates and each consistent with the same $\Delta(\vec{k})$ at that
doping. Thus Bogoliubov-like quasi-particles appear to exist at
these momentum space locations at all dopings. This is consistent
with the small motion of the FS in this doping range detected by
ARPES.

These observations certainly do not exhaust the changes observed
in $g(\vec{r},E)$ with falling doping. A very strong effect is the
evolution, with doping, of the $\vec{q}$-space location of
strongest $LDOS$-modulation at any energy. This modulation is
always associated with $\vec{q}_1$ and vits location evolves from
$\vec{q}_1=\frac{2\pi}{6a_0}$ at p$=0.19\pm0.01$, to
$\vec{q}_1=\frac{2\pi}{5.1a_0}$ at p$=0.14\pm0.01$, to
$\vec{q}_1=\frac{2\pi}{4.7a_0}$ at p$=0.10\pm0.01$. Another effect
is a decrease in relative intensity of dispersive
$LDOS$-modulations $\vec{q}_2$, $\vec{q}_3$, $\vec{q}_6$,
$\vec{q}_7$ relative to those of  $\vec{q}_1$, $\vec{q}_5$, with
falling with p.  These effects will be reported in detail
elsewhere.

The doping dependence of states with $\vec{k}=(\pm\pi/a_0,0)$ and
$(0,\pm\pi/a_0)$ in the 'flat band' region near the
zone-face\cite{damascelli03} (green shaded areas in Fig.5{\bf A})
is extremely different. These states can also be identified by
FT-STS analysis of $g(\vec{r},E)$ data. By definition, the
coherence peaks in $g(\vec{r},E)$ occur at $E=\Delta(\vec{r})$. In
all samples, they exhibit intense particle-hole symmetric
$LDOS$-modulations, with wavevectors $\vec{G}=(\pm 2\pi/a_0,0)$
and $(0,\pm 2 \pi/a_0)$.\cite{mcelroy03} These coherence peak
$LDOS$-modulations at $E=\Delta$ possibly occur due to Umpklapp
scattering between $\vec{k}=(\pm\pi/a_0,0)$ and
$(0,\pm\pi/a_0)$\cite{mcelroy03}. Therefore, the coherence peaks
in tunneling are identified empirically with the zone-face states
at $\vec{k}=(\pm\pi/a_0,0)$ and $(0,\pm\pi/a_0)$. This
identification is also consistent with theory.  The coherence
peaked tunneling spectra (e.g. Fig.2{\bf F}: spectra 1-4) are
theoretically viewed as due to superconducting pairing on the
whole FS because such spectra are consistent with a
$\Delta_{x^2-y^2}$ everywhere on the ARPES-determined FS near
optimal doping.\cite{norman03} We therefore consider any spatial
regions that show clear coherence peaks with $\vec{q}=\vec{G}$
$LDOS$-modulations, to be occupied by a canonical d-wave
superconductor (dSC).

Near optimal doping, more than 98\% of any FOV exhibits this type
of coherence peaked dSC spectrum. As the range of local values of
$\Delta(\vec{r})$ increases with decreasing doping, the intensity
of the $\vec{q}=\vec{G}$ coherence peak $LDOS$-modulations becomes
steadily weaker until, wherever $\Delta(\vec{r})>65$ meV, they
disappear altogether. This process can be seen clearly in the
gap-averaged spectra of Fig.3{\bf F} where the average height of
the coherence peaks declines steadily with increasing $\Delta$. It
is found equally true for all dopings. Wherever the coherence
peaks and their $\vec{q}=\vec{G}$ $LDOS$-modulations are absent, a
well-defined new type of spectrum is always observed. Figure 6{\bf
A} shows a high resolution gapmap from a strongly underdoped
sample. Examples of this new type of spectrum, along with those of
coherence peaked dSC spectra, are shown in Fig.6{\bf B}. The
coherence peaked spectra (red) are manifestly distinct from the
novel spectra (black) which have a V-shaped gap reaching up to
-300 meV and +75 meV. For reasons to be discussed below, we refer
to the new spectrum as the zero temperature pseudogap (ZTPG)
spectrum.

The replacement of coherence peaked dSC spectra by ZTPG spectra
first begins to have strong impact on averaged properties of
$g(\vec{r},E)$ and $g(\vec{q},E)$ below about p=0.14 where the
fractional area covered by ZTPG spectra first exceeds ~10\% of the
FOV. In terms of the spectral shape no further evolution in the
form of the ZTPG spectrum is detected at lower dopings. Instead, a
steadily increasing fractional coverage of the surface by these
ZTPG spectra is observed. Our previous studies\cite{pan01,lang02}
were carried out at dopings $p>0.14$ and, when ZTPG spectra have
been detected at such higher dopings,\cite{pan01,howald01} it is
in a tiny fraction of the FOV. Very significantly, spectra similar
to the ZTPG spectrum are detected inside cores of Bi-2212
quantized vortices where superconductivity is
destroyed.\cite{renner98,pan00} Furthermore, a very similar
spectrum is observed in another very underdoped cuprate
Na$_x$Ca$_{2-x}$CuO$_2$Cl$_2$,\cite{koshaka03,hanaguri04} even in
the non-superconducting phase. It therefore seems reasonable that
the characteristic zero-temperature spectrum in the pseudogap
phase is of this type. This is why we tentatively assign this
spectrum the ZTPG designation.

As discussed above, the $g(\vec{r},E)$ for $E<\bar{\Delta}/2$ in
even the most underdoped samples (Fig.3{\bf E}, Fig.4{\bf C})
exhibit relatively homogenous electronic structure with good
quasi-particles dispersing on the Fermi-arc. However, for
$E>\bar{\Delta}/2$ in these same samples, our previous analyses
techniques fail, probably because very different phenomena are
occurring in different nanoscale regions of each FOV. To explore
the implications of the ZTPG spectrum for strongly underdoped
samples, new analysis techniques are therefore required.  Here we
introduce a masking process which has proven highly effective.
From a given strongly underdoped data set, the $g(\vec{r},E)$ in
all regions where $E\Delta>65$ meV are excised and used to form a
new masked data set $g(\vec{r},E)|_{\Delta>65}$. The remainder
forms a second new dataset $g(\vec{r},E)|_{\Delta<65}$. The
$E\Delta>65$ meV cutoff was chosen because, on the average, it
represents where the coherence peaks with associated
$\vec{q}=\vec{G}$ modulations have all disappeared and are
replaced by the ZTPG spectra. An example of this type of mask for
the gapmap in Fig.6{\bf A} is show in Fig.5{\bf C}. It is
important to note a serious drawback of the masking process. The
$\vec{q}$ resolution is of masked data is considerably worse in
than those shown in Fig.5{\bf B} because the largest contiguous
nano region in the mask is about 20\% of the full FOV. As a
result, the precise modulation period and dispersions of any
effects detected by masking cannot can be determined with nearly
the same degree of accuracy as the low-energy quasi-particle
interference signal (Fig.5 and Ref. 28).

FT-STS analysis of such ($g(\vec{r},E)|_{\Delta<65}$,
$g(\vec{r},E)|_{\Delta>65}$) pairs shows that they exhibit
dramatically different phenomena. In the
$g(\vec{r},E)|_{\Delta<65}$, the dispersive trajectory of
$\vec{q}_1$ is seen up to $E \approx 36$ meV and no further
$LDOS$-modulations can be detected at any higher energy (red
symbols in Fig.6{\bf D}). In the $g(\vec{r},E)|_{\Delta>65}$ data,
the identical dispersive $\vec{q}_1$ signal is seen below E~36
meV. However, a new $LDOS$-modulation appears in the
$g(\vec{r},E)|_{\Delta>65}$ between E$>65$ meV and our maximum
energy E=150 meV (black symbols in Fig.6{\bf D}). We designate its
wavevector $\vec{q}^*$.

To explore the real-space structure of this new high-energy
$LDOS$-modulation, we define a map
$\Gamma_{65}^{150}(\vec{r})=\Sigma_{E=65}^{150}g(\vec{r},E)|_{\delta>65}$
which sums over this energy range. This map is shown in Fig.7{\bf
A} and, although it has quite a disordered mask, careful
examination reveals checkerboard modulations within each nano
region. Importantly, Fourier transform analysis of this
$\Gamma_{65}^{150}(\vec{r})$ shown in Fig. 7B reveals a
well-defined wavevector set $\vec{q}^*=(\pm 2 \pi/4.5a_0,0)$ and
$(0,\pm 2 \pi/4.5a_0)\pm15\%$ for these new high-energy
modulations as indicated by the arrow on black data points in
inset to Fig. 7B. The identical analysis the complementary map
$\Sigma_{65}^{150}g(\vec{r},E)|_{\delta<65}$ is featureless near
$\vec{q}^*$ (red data in inset of Fig. 7B). Thus, an
$LDOS$-modulation with very low (or zero) dispersion, exists above
$\pm 65$ meV exclusively in regions characterized by ZTPG spectra
of strongly underdoped Bi-2212 samples.

Constant-current topography represents, albeit logarithmically,
the contour of constant integrated density of states up to the
sample-bias energy. It does not suffer from the systematic
problems due to effects of the constant-current setup condition
renormalization\cite{pan01} which plague $g(\vec{r},E)$. It is
therefore a more conclusive technique for detection of net charge
density modulations by STM. To search for topographic modulations,
we apply the identical mask (Fig.7{\bf C}) to the topographic
image which was acquired simultaneously with the gapmap in 6{\bf
A}. The magnitude of the Fourier transform along the
$\vec{q}||(2\pi,0)$ for the masked topographic image shows that,
in the ZTPG  regions, the topography is modulated with
$\vec{q}_{topo}=(\pm 2 \pi/5a_0,0)$ and $(0,\pm 2 \pi/5a_0)\pm
25\%$(indicated by the arrows in Fig.7{\bf D}). Fourier transforms
of the complementary part of the topography (from $\Delta<$65 meV
regions not exhibiting ZBTG spectra) shows no such modulations at
any wavelengths near this $\vec{q}_{topo}$ (red in Fig.7{\bf C}).
No topographic modulations near $\vec{q}_{topo}$ are detected
anywhere in samples at higher doping. An additional peak near
$\vec{q}_{recon}=(2 \pi/2a_0,\pm 2 \pi/2a_0)$ and $(-2 \pi/2a_0,-
2 \pi/2a_0)$ in the Fourier transform of the topograph comes from
a reconstruction along the supermodulation maximum, is observed at
all dopings and, since no signature is observed in $LDOS$ at
$\vec{q}_{recon}$ for any energy or doping, is regarded as
irrelevant.

Some important new facts about the strongly underdoped regime of
Bi-2212 electronic structure emerge from these results. Our first
finding is related to the Fermi-arc quasi-particle states. As
shown in Fig. 5C, FT-STS indicates that quasi-particle
interference occurs between Bogoliubov-like states in
approximately the same region of $\vec{k}$-space for all dopings.
These Fermi-arc quasiparticles remain spatially homogenous (except
for relatively weak $LDOS$-modulations) even in the most
underdoped samples. They are Bogoliubov-like in the sense that
they exhibit particle-hole symmetry at each location in
$\vec{k}$-space, and, at each doping, are all consistent with the
same $\Delta(\vec{k})$. Therefore, one can reasonably postulate
that Fermi-arc states are gapped by superconducting interactions
at all dopings studied. If so, they, and the associated 'nodal'
superconducting
state,\cite{senthil00,vojta00,joglekar03,parcollet03} are
amazingly robust against the heterogeneous electronic phenomena
which so dominate Bi-2212 at other energies.

Our second finding is the very different fate of states in the
flat-band regions near
$\vec{k}\approx(\pm\pi/a_0,0);(0,\pm\pi/a_0)$. The appearance of
ZTPG spectra in strongly underdoped samples coincides exactly with
destruction of antinodal superconducting coherence peaks.
Exclusively in these ZTPG regions (black in the gapmap), we find
the three new modulation phenomena: (1) topographic modulations
with $\vec{q}_{topo}=(\pm 2 \pi/5a_0,0)$ and $(0,\pm 2
\pi/5a_0)\pm 25\%$, (2) a peak in $LDOS$ centered around
$\vec{q}^*=(\pm 2 \pi/4.5a_0,0)$ and $(0,\pm 2 \pi/4.5a_0)\pm
15\%$ for $E>$65 meV to at least E=150 meV and, (3) the dispersive
$\vec{q}_1$ quasi-particle branch exhibits its maximum modulation
intensity when it passes through
$\vec{q}_1=\vec{q}_{topo}=\vec{q}^*$ (Fig 7D). This last situation
has been
predicted\cite{zhu01,chen03yeh,wang03,capriotti03,podolsky03} as a
consequence of the Fermi surface geometry and quasi-particle
dispersion  in the presence of potential scattering from
charge-order with fixed $\vec{Q}$; in this case
$\vec{Q}=\vec{q}_{topo}=\vec{q}^*$. Taken together, these
observations all point to the appearance of an unusual charge
ordered state with $\vec{q}^*=(\pm 2 \pi/4.5a_0,0)$ and $(0,\pm 2
\pi/4.5a_0)\pm 15\%$, occurring only in the regions characterized
by the ZTPG spectrum and only in strongly underdoped Bi-2212. Note
that due to the strong disorder and the limited size of the mask
domain we can not distinguish between a charge density modulation
caused by a condensed charge order and that caused by impurity
through an enhanced charge susceptibility in a state without
condensed charge order.\cite{fu04} In addition we do not imply
that there is true charge long range order since from the Imry-Ma
argument this is always absent in a disordered system like
Bi-2212.

Charge order has been observed in other underdoped cuprates,
including Nd doped La-214 with inelastic neutron
scattering\cite{tranquada95} and more recently by STM in
Na$_x$Ca$_{2-x}$CuO$_2$Cl$_2$.\cite{hanaguri04} It has also been
proposed, based on reported nondispersive (between 0 and 20 meV),
'line object', $LDOS$ modulations with
$\vec{q}=(2\pi/4a_0,0)$\cite{howald03}, that static stripes exist
in optimally doped Bi-2212 below T$_c$. However, none of these
phenomena have been detected in several independent higher
resolution studies.\cite{hoffman02q,mcelroy03,vershinin04}

Other very suggestive findings have also been made by STM in
Bi-2212. Field induced sub-gap $LDOS$-modulations, with
$\vec{q}_{vortex}=(\pm 2 \pi/4.3a_0,0)$ and $(0,\pm 2
\pi/4.3a_0)\pm 15\%$ were discovered surrounding vortex cores
(where superconductivity is destroyed) near optimal doping. This
observation provided the first STM evidence for some type of
incipient charge-order competing with superconductivity in
cuprates.\cite{hoffman02v} Pioneering STM experiments to map the
low energy $LDOS$ above T$_c$ when superconductivity is also
destroyed, have detected sub-gap LDOS-modulations with
$\vec{q}_{PG}=(\pm 2 \pi/4.6a_0,0)$ and $(0,\pm 2
\pi/4.6a_0)$\cite{vershinin04} at near-optimal doping. Although
neither of these low energy phenomena are fully understood, they
do, along with low energy phenomena in ZTPG regions reported here,
form a triad of apparently consistent observations. Destruction of
superconductivity, whether by high magnetic fields, by exceeding
T$_c$, or by strong underdoping, results in very similar effects
on low energy $LDOS$ modulations. It remains to be determined how
the vortex core and pseudogap observations relate directly to the
charge order.

The identity of the electronic phase represented by the
$\vec{q}_{topo}=\vec{q}^*$ charge order is difficult to discern.
In the absence of disorder, plaquette orbital-order phases such as
staggered flux phase (SFP)\cite{affleck88,lee03} and
D-density-wave (DDW),\cite{chakravarty01} or intra-plaquette
orbital phases,\cite{varma97} are not expected to exhibit
topographic or $LDOS$-modulations. However, in theory,
$LDOS$-modulations can be produced by vortex and disorder
scattering in the SFP and DDW phases but not near $\vec{q}=(\pm 2
\pi/4a_0,0)$ and $(0,\pm 2
\pi/4a_0)$.\cite{peregbarnea03,bena03,kishine01} It remains
theoretically unexplored whether disordered orbital phases could
result in the complete set of new phenomena (see below) we report
here. Charge-ordered phases including
stripes,\cite{zaanen89,kivelson98,low94} disorder pinned
electronic liquid crystal,\cite{kivelson03} strong-coupling spin-
and charge-density waves,\cite{sachdev03} and, recently, hole-pair
crystals\cite{chen03zhang} have been proposed to exist in
underdoped cuprates. Each of these phenomena would yield both
$LDOS$- and topographic-modulations. But again, it remains
theoretically unexplored whether these theories can account for
our observations that in the ZTPG regions (i) a characteristic new
tunneling spectrum exists, (ii) the $\vec{q}^*$ modulations appear
only above a relative high energy ($\approx 65$ meV), (iii) they
exhibit an incommensurate wavevector $\vec{q}=(\pm 2
\pi/4.5a_0,0)$ and $(0,\pm 2 \pi/4.5a_0)\pm15\%$, (iv) they
exhibit the same spatial phase for positive and negative biases -
so that the filled-state density maxima coincide with the
empty-state-density maxima, and (iv) the $\vec{q}^*$ modulations
are replaced by the dispersive quasi-particle interference signals
at sub-gap energies. A further point is that the predicted strong
breaking of the $90^{\circ}$ rotation symmetry in the stripe
scenario is not observed in any of the STM studies, but again, it
may be possible that this is due to the presence of strong
disorder. For all these reasons the precise identity of the
charge-order state in strongly underdoping Bi-2212 remains
elusive.

The data reported here also motivate a new conjecture on the
evolution of electronic structure with reduced doping in Bi-2212.
In  $\vec{r}$-space, we identify two extreme types of $LDOS$
spectra (Fig. 6B). The first exhibits clear coherence peaks at the
gap edge and dominates near-optimal samples. The second type
(ZTPG-spectrum) exhibits a V-shaped gap over a much wider energy
range and dominates in strongly underdoped samples.  We associate
the former with a pure d-wave superconducting state and conjecture
that the latter reflects a zero temperature charge ordered state
existing at sufficiently low dopings in the pseudogap regime. If
the whole Bi-2212 sample were homogeneous and consisted of only
one of the above phases, then, in  $\vec{k}$-space, quasi-particle
peaks would exist all along the Fermi surface in the pure d-SC
phase, but only on a finite arc around the gap nodes in the charge
ordered phase with the zone face states being incoherent due to
localization. This may indeed be the case in
Na$_x$Ca$_{2-x}$CuO$_2$Cl$_2$.\cite{hanaguri04,ronning03} In
Bi-2212 he reality is more complicated. In optimally doped
samples, more than 98\% of the surface area exhibit dSC spectra
(Fig. 3A,B). As doping falls the ZTPG regions appear and grow in
significance until as doping approaches p~0.1, almost 60\% of the
area exhibits the ZTPG characteristics and the associated
charge-order (Fig. 3E). From this trend, it is reasonable to
expect that, at even lower doping in the zero temperature
pseudogap phase, 100\% of the sample would exhibit the ZTPG
characteristics and be charge-ordered.

In a spatially disordered situation, the probability of occurrence
of the two types of phenomena evolves continuously with doping in
a fashion related to the evolution of the gapmaps in Fig. 3.
Therefore, properties which average over nanoscale phenomena would
appear to evolve smoothly between the two extremes. Experimental
results which are relevant to this proposal include, for example,
the doping dependence of the  ARPES $(\pi,0)$
peak,\cite{timusk99,damascelli03,campuzano02,feng00} the specific
heat jump at the superconducting phase transition,\cite{timusk99}
and the c-axis conductivity.\cite{timusk99} Due to the
heterogeneous mixture of the dSC and the ZTPG regions, it is
difficult for a spatially averaged experiment like ARPES to
discern the properties of the ZTPG region. However, the well-known
tendency of the coherent quasiparticle peaks near the zone face to
be suppressed by
underdoping,\cite{damascelli03,campuzano02,feng00} is consistent
with our conjecture. The specific heat jump at the superconducting
transition is a characteristic specifically of a dSC phase and not
of the charge-ordered ZTPG phase. Hence, the declining specific
heat jump as a function of underdoping\cite{timusk99,loram01} is
also consistent with our hypothesis. Finally, due to tunneling
matrix element effects, c-axis tunneling senses the zone-face
quasiparticles instead of the nodal ones. Hence the decrease of
the c-axis conductivity with underdoping\cite{timusk99} also seems
consistent with our suggestion. Further inter-comparison between
the results of these experimental techniques will be required to
explore these proposals.

One may also wonder if the charge-ordered ZTPG phase could
actually be the source of the high temperature pseudogap
phenomena.\cite{timusk99} Since the charge-order would melt as a
function increasing temperature, the answer to this question it is
not clear. If the charge order melting temperature were actually
the pseudogap temperature T$^*$, then of course, the entire
pseudogap phenomena would be caused by the ZTPG phase. On the
other hand if the charge melting temperature were lower than the
pseudogap temperature T$^*$, then there would exist yet another
mechanism that gaps the spin excitations at the pseudogap
crossover.

Independent of these conjectures on doping-dependence of
spatially-averaged observables, the data reported here represent
two significant advances. First, a charge ordered state competing
with superconductivity exists in strongly underdoped  Bi-2212 and
second, the charge-ordered regions and the d-SC regions share the
same low energy nodal quasiparticles. New microscopic theories for
the electronic structure of underdoped cuprates are therefore
required to explain, not only the unusual characteristics of the
charge order phenomenon and how it competes with superconductivity
for spectral density at the zone-face, but also how it meshes so
smoothly with the nodal superconductivity at low energies.

{\bf Figure 1}\\
A schematic phase diagram of the cuprates. Note
that the region of our studies is between $0.1<p<0.2$ and near the
zero temperature axis of this diagram where the magnetically
disordered spin glass region coexists with the superconductivity.

{\bf Figure 2}\\
{\bf A} The gapmap $\Delta(\vec{r})$ calculated from a
$g(\vec{r},E)$ data set which was measured in a 55 nm field of
view (FOV) with $256^2$ pixels; inset topography in this FOV. {\bf
B} The unprocessed linecut $g(\vec{r},E)$ connection a region with
sharp coherence peaks with a region designated ZTPG (see text),
along the red line in Fig. 2{\bf A}. {\bf C} The measured
$g(\vec{r},E=-12$meV$)$ in the identical FOV as 2{\bf A}. {\bf D}
calculated from {\bf C} (the reciprocal space location of the Bi
or Cu atoms are labeled $(2\pi,0)$. These five types of
measurements are used in concert to study the doping dependence of
electronic structure in Bi-2212.

{\bf Figure 3}\\
{\bf A-E} Measured $\Delta(\vec{r})$ for five different
hole-doping levels. {\bf F}. The average spectrum associated with
each gap value in a given FOV. They were extracted from the
$g(\vec{r},E)$ that yielded Fig. 3C but the equivalent analysis
for $g(\vec{r},E)$ at all dopings yields results which are
indistinguishable.  The coherence peaks can be detected in \#'s
1-4.

{\bf Figure 4}\\
{\bf A-C} Examples of measured $g(\vec{q},E)$ for a variety of
energies $E$ as shown at three doping levels.

{\bf Figure 5}\\
{\bf A.} A schematic representation of the 1st Brillouin zone and
Fermi surface location of Bi-2212. The flat-band regions near the
zone face are shaded in blue. The eight locations which determine
the scattering within the octet model are show as red circles and
the scattering vectors which connect these locations  are show as
arrows labeled by the designation of each scattering vector.  {\bf
B}. Measured dispersions of the $LDOS$-modulations $\vec{q}_1$,
$\vec{q}_5$ and $\vec{q}_7$ for the 3 dopings whose unprocessed
data is shown in Fig.4. We chooses this set of three
$\vec{q}$-vectors because they exhibit the maximum intensity of
any set sufficient to independently determine the locus of
scattering $\vec{k}_s(E)$ for all dopings. {\bf C}. Calculated
loci of scattering $\vec{k}_s(E)$ for all 3 dopings. The blue line
is a fit to the 89KOD data

{\bf Figure 6}\\
{\bf A} A high-resolution ~46nm square gapmap from a strongly
underdoped sample. {\bf B}. Examples of representative spectra
from (1) nanoscale regions exhibiting coherence peaked spectra
with $\vec{q}=\vec{Q}$ $LDOS$-modulations at $E=\Delta(\vec{r})$
(red) and (2) from regions exhibiting ZTPG spectra (black). The
locations where these spectra occur are shows as small dots in
{\bf A.} {\bf C}.  The mask identifying regions with $\Delta<$65
meV from the ZTPG regions calculated from gapmap in {\bf A}. {\bf
D}. Dispersion of $\vec{q}_1(E)$ in regions with coherence peaked
spectra $\Delta < 65$ meV is shown in red. There are no
modulations at any higher energies within our range. Dispersion of
$\vec{q}_1(E)$ in regions with ZTPG spectra for E$< 36$ meV (black
symbols). The red squares in Fig.5{\bf B} represent a combination
of these two (indistinguishable) dispersions in this range. For
E$>65$ meV, the wavevector of the  new modulations in ZTPG regions
are shown in black. To within our uncertainty they do not disperse
and exhibit $\vec{q}^*=(\pm 2 \pi/4.5a_0,0)$ and $(0,\pm 2
\pi/4.5a_0)\pm15\%$.

{\bf Figure 7}\\
{\bf A}The image of $g(\vec{r},E)$ masked by Fig.6{\bf C} and then
summed from 65 meV to 150 meV
:$\Gamma_{65}^{150}(\vec{r})=\Sigma_{65}^{150}g(\vec{r},E)|_{\delta>65}$.
The light grey regions are outside the mask.  This FOV is ~46nm
square and a careful examination reveals a 'checkerboard'
modulation occurring within the regions inside the mask. The
equivalent image for
$\Gamma_{65}^{150}(\vec{r})=\Sigma_{65}^{150}g(\vec{r},E)|_{\delta<65}$
is featureless except for Bi atom locations, and the
supermodulation which is at 45$^{\circ}$ to the modulation in {\bf
A}. {\bf B}. Fourier transform {\bf A}. The Bi atom locations are
circled in orange. A square of diffuse maxima are observed
surrounding the central point. The inset shows the plot of the
Fourier transform amplitude along the line shown. It reveals a
maximum at $\vec{q}^*=(\pm 2 \pi/4.5a_0,0)$ and $(0,\pm 2
\pi/4.5a_0)\pm15\%$ as indicated by the arrow. {\bf C}. The
magnitude of the Fourier transform of the masked topographic
image, taken at 150 mV and 150 pA, along the
$\vec{q}_{topo}||(2\pi,0)$ direction is modulated with
$\vec{q}_{topo}=(\pm 2 \pi/5a_0,0)$ and $(0,\pm 2 \pi/5a_0)\pm
25\%$ (black squares the black line is a guide to the eye).
Fourier transforms of the complementary part of the topography
(from $<65$ meV regions not exhibiting ZBTG spectra) shows no such
modulations at any wavelengths near this $\vec{q}_{topo}$ (red
triangles the red line is a guide to the eye). The difference
between these two Fourier transform intensities is shown in blue
and shows the degree to which the ZTPG region shows topographic
modulations undetectable elsewhere. The relative weakness of the
topographic modulations at $\vec{q}_{topo}$ is because of the
logarithmic sensitivity of constant-current topography to net
charge modulation. D. A plot of the amplitude of the $\vec{q}_1$
$LDOS$ modulation as a function of $|\vec{q}_1|$ for the sample
data as in Fig.3{\bf C}, 4{\bf E}, 5{\bf D}. The maximum intensity
of the modulations in the ZTPG regions occurs at
$|\vec{q}_1|=2\pi/4.8a_0±10\%$. No special scattering is observed
of the quasiparticles in the dSC regions.

\begin{acknowledgments}
We acknowledge and thank A. V. Balatsky, S. Chakravarty, M. P. A.
Fisher, T. Hanaguri, S. Kivelson, P. A. Lee, A. Millis, D. Pines,
S. Sachdev, J. Sethna, T. Uemura, A. Yazdani, J. Zaanen, and S.-C.
Zhang  for very helpful discussions and communications.
\end{acknowledgments}

\bibliography{to_post_3_18_04}

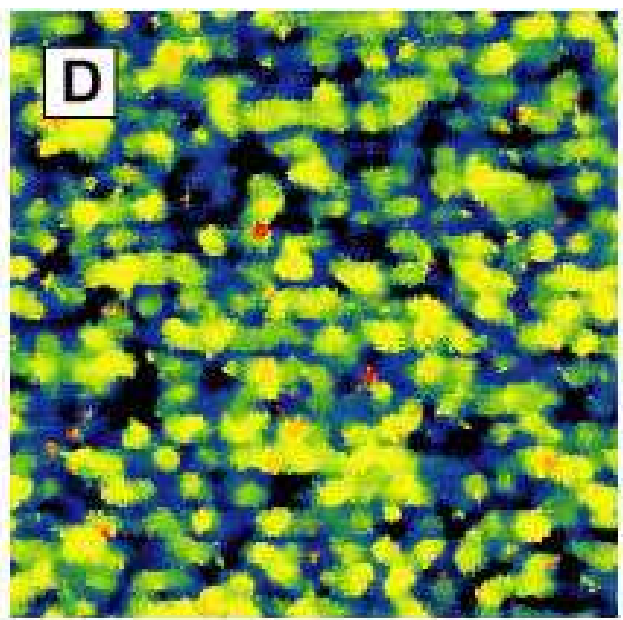
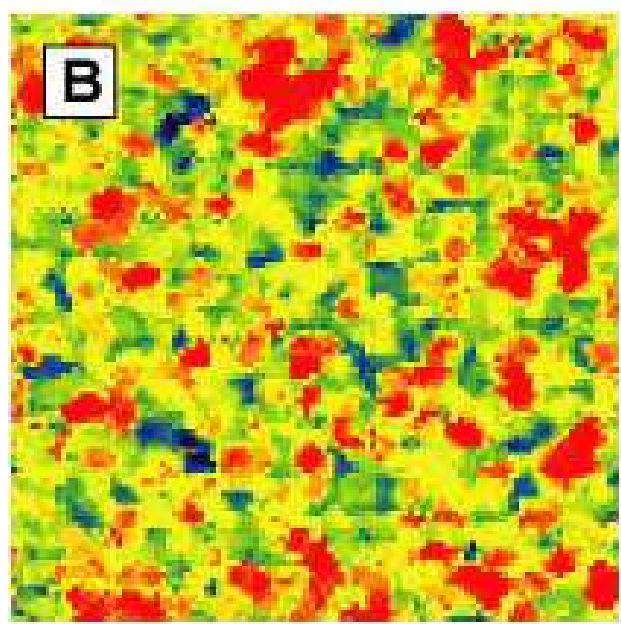
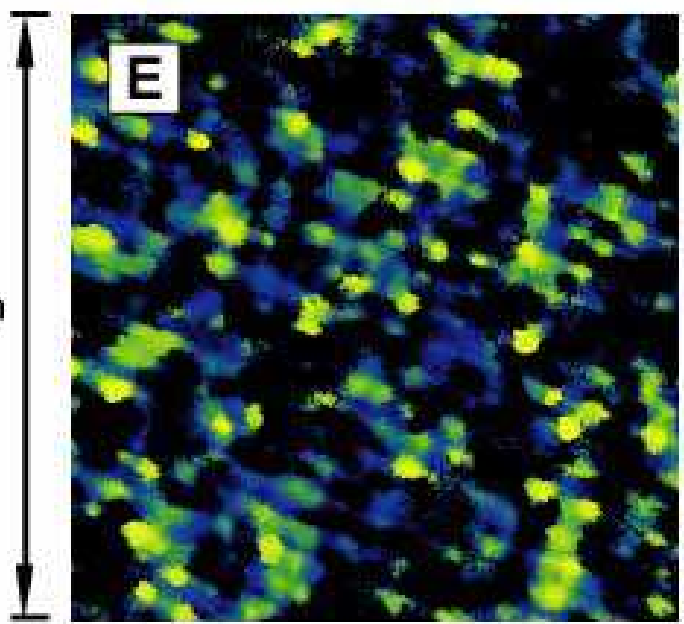
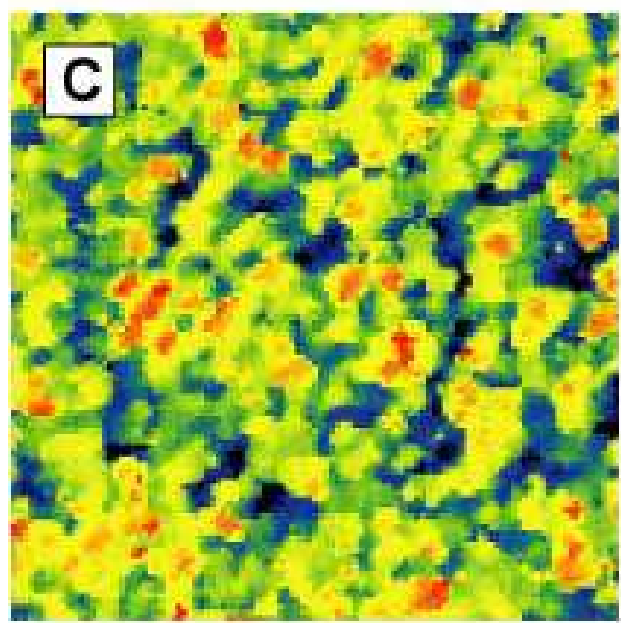
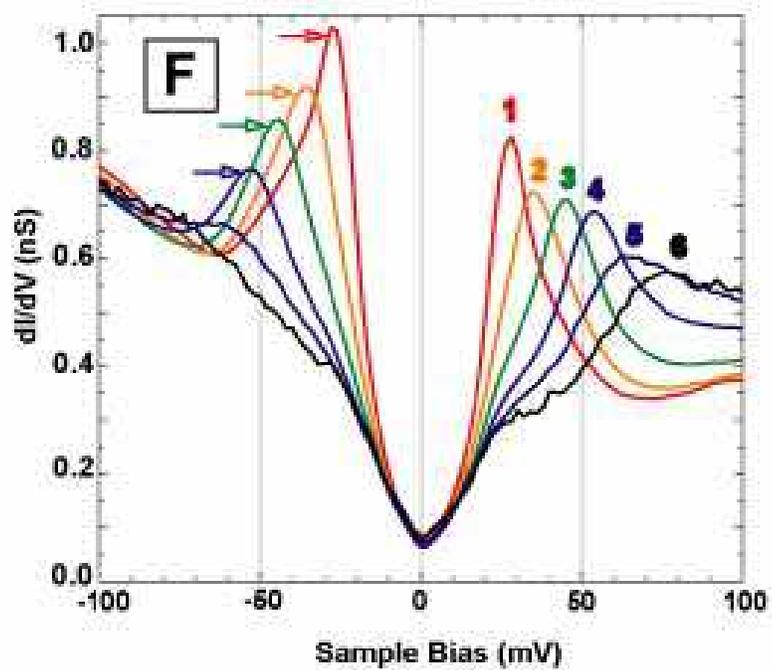
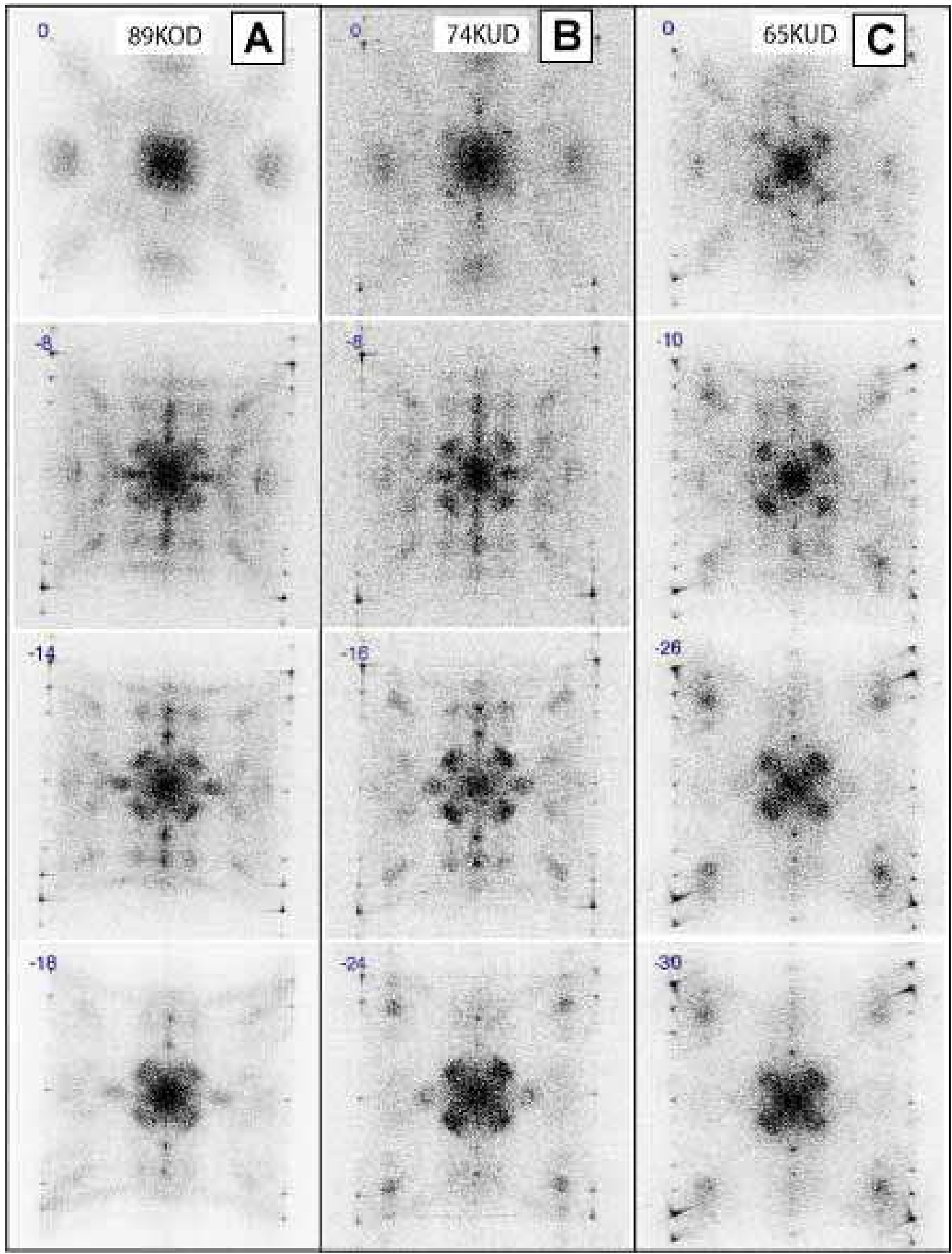
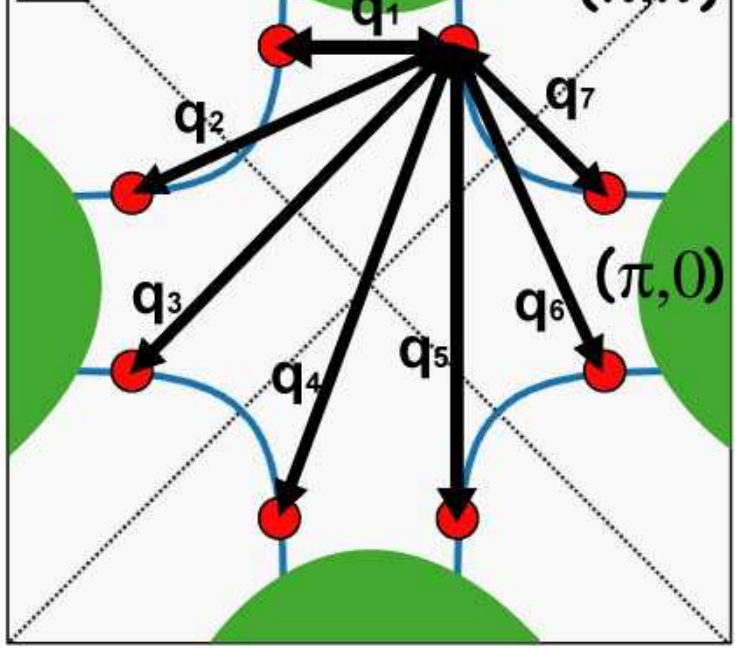
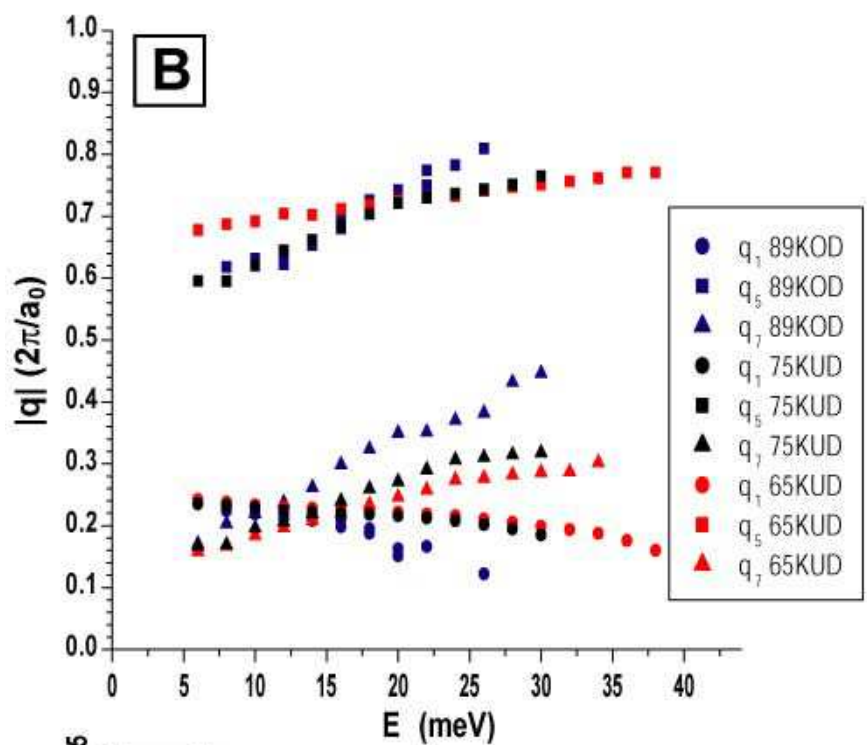
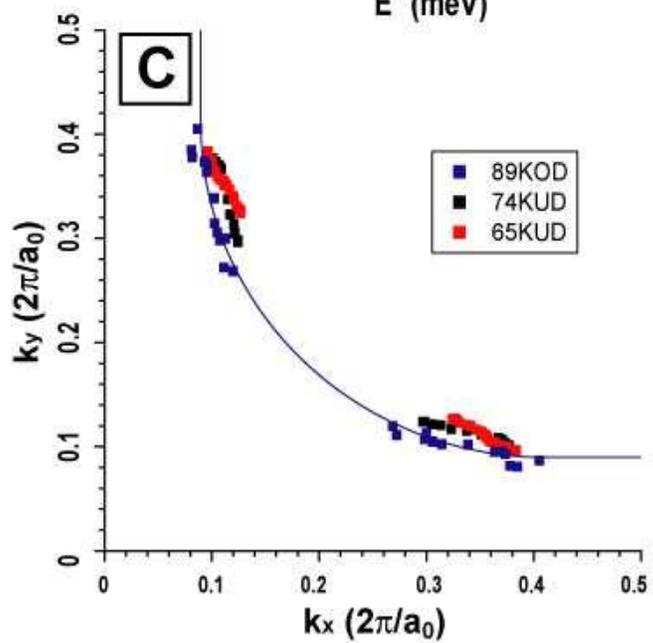
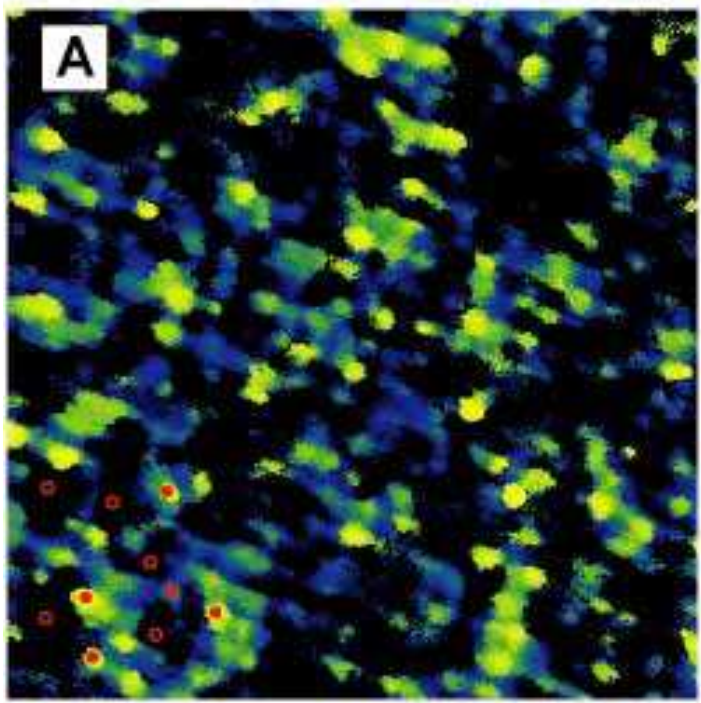
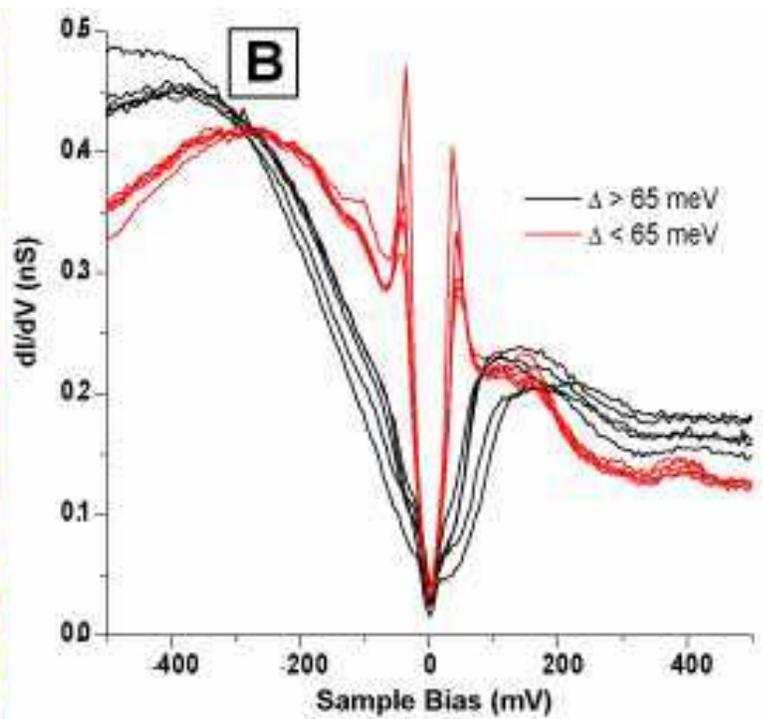
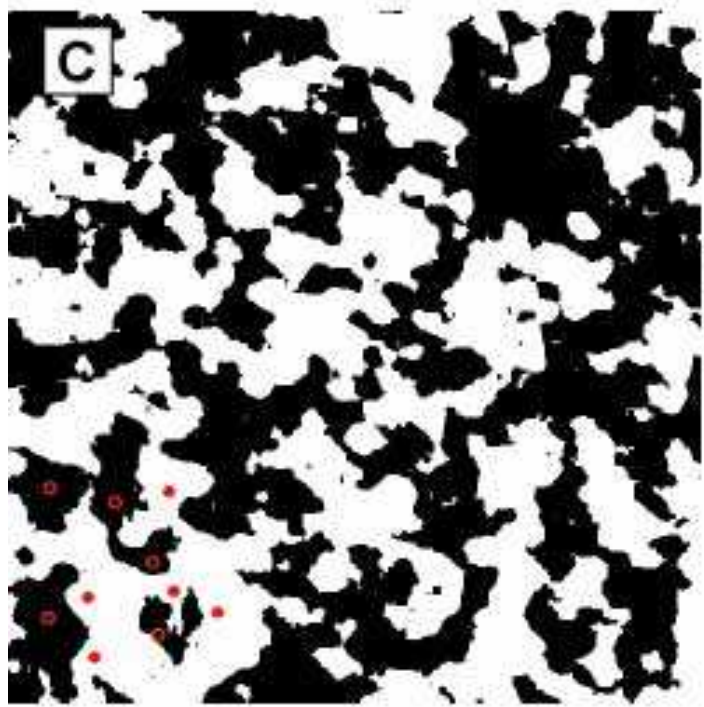
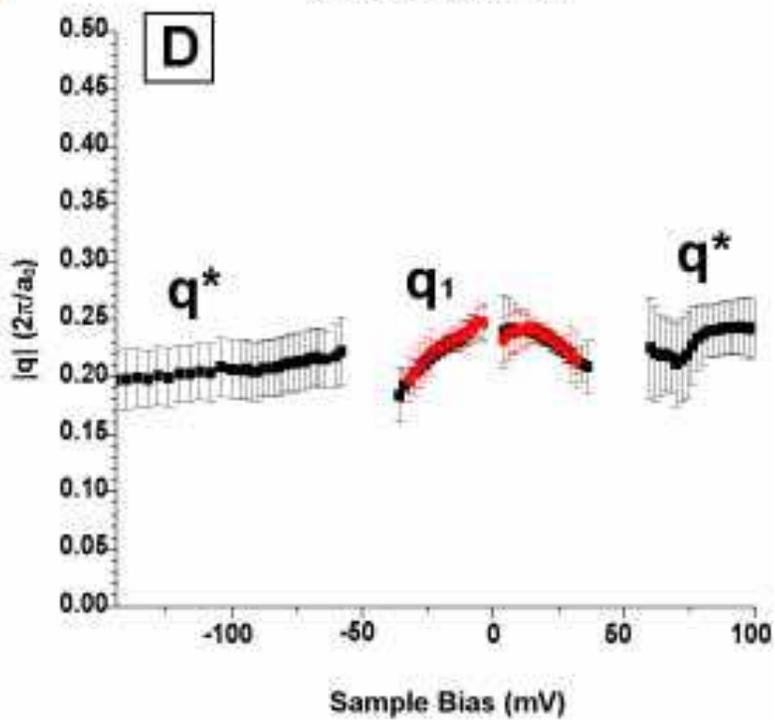
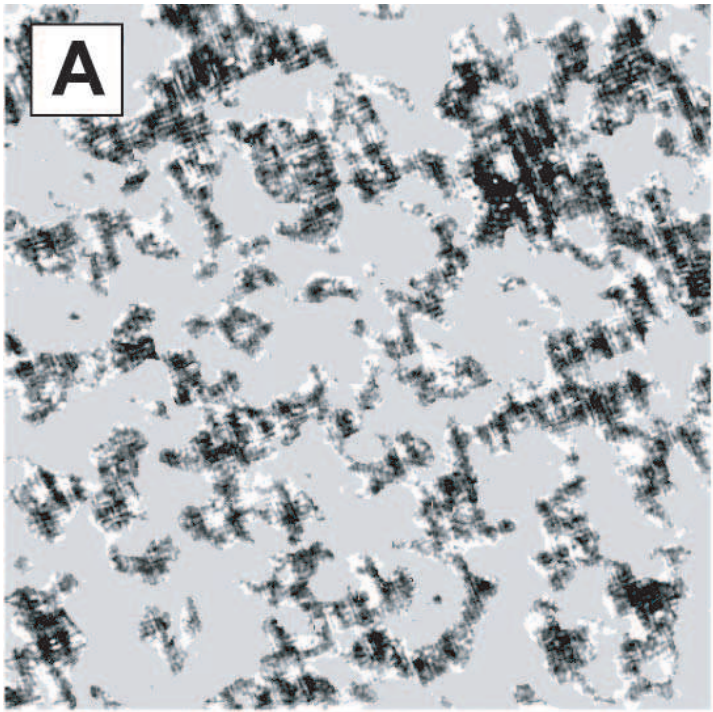
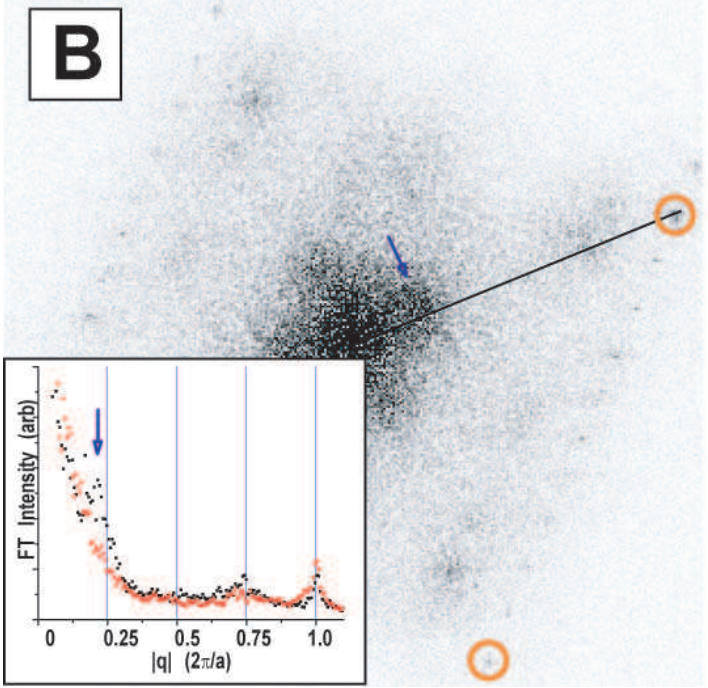
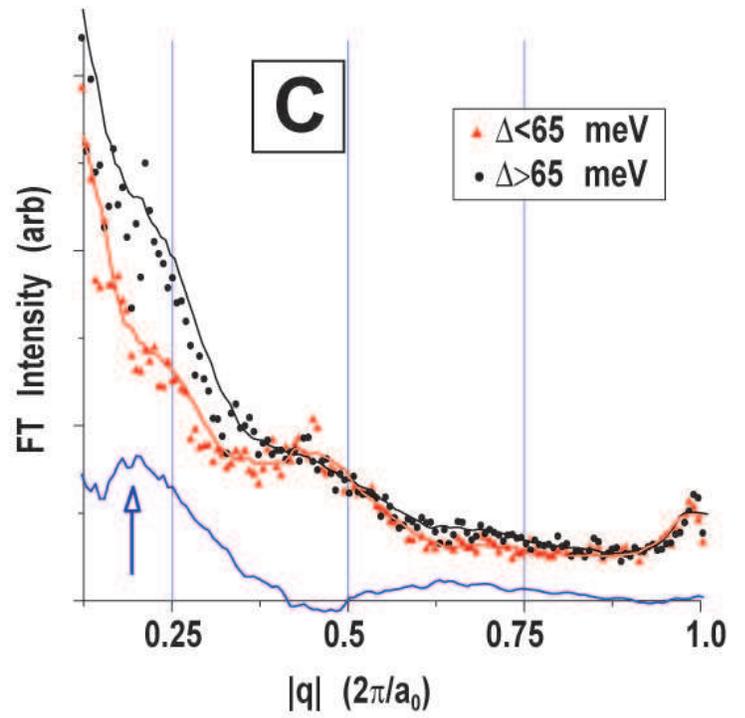
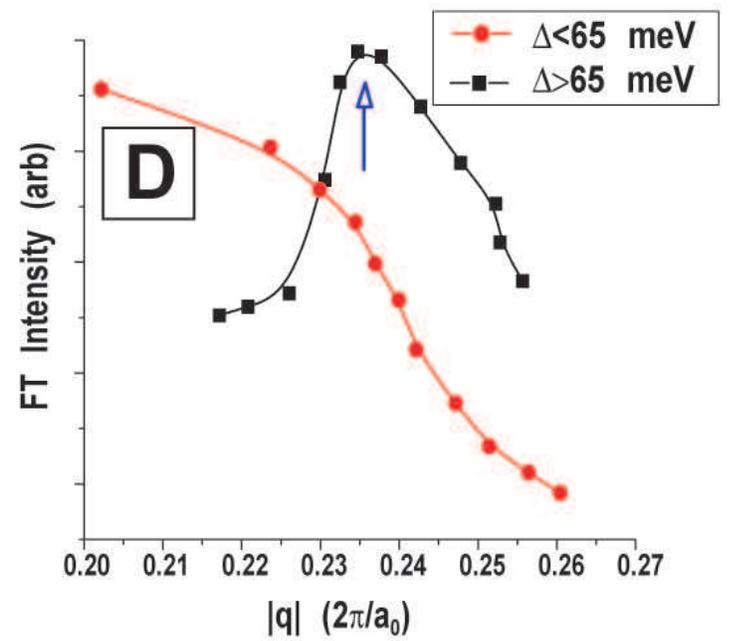
\end{document}